\documentclass[
aps,
reprint,
superscriptaddress,
noeprint,
amsmath,amssymb,
prb,
showkeys,
floatfix,nofootinbib
]{revtex4-2}

\usepackage[dvipsnames]{xcolor}
\usepackage{graphicx}
\usepackage[hidelinks=true,colorlinks=true,linkcolor=blue,citecolor=blue]{hyperref}
\usepackage[utf8]{inputenc}
\usepackage[T1]{fontenc}
\usepackage{physics}
\usepackage{mathtools}
\usepackage{siunitx}
\usepackage{bm}
\usepackage{dcolumn}
\usepackage{appendix}
\usepackage{lipsum}
\usepackage{multirow}
\usepackage{placeins}
\usepackage{soul}
\newcommand{\orcid}[1]{\href{https://orcid.org/#1}{\includegraphics[width=8pt]{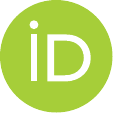}}}
\usepackage[normalem]{ulem}

\begin{document}
\title{Covariant Jacobi-Legendre expansion for total energy calculations within the projector-augmented-wave formalism}

\author{Bruno Focassio\orcid{0000-0003-4811-7729}}\email{bruno.focassio@lnnano.cnpem.br}
\affiliation{Brazilian Nanotechnology National Laboratory (LNNano/CNPEM), 13083-100, Campinas, SP, Brazil}

\author{Michelangelo Domina\orcid{0000-0002-7009-9240}}
\affiliation{School of Physics and CRANN Institute, Trinity College Dublin, Dublin 2, Ireland}

\author{Urvesh Patil\orcid{0000-0002-3924-653X}}
\affiliation{School of Physics and CRANN Institute, Trinity College Dublin, Dublin 2, Ireland}

\author{Adalberto Fazzio\orcid{0000-0001-5384-7676}}
\affiliation{Ilum School of Science, CNPEM, 13083-970 Campinas, São Paulo, Brazil}
\affiliation{Federal University of ABC (UFABC), 09210-580 Santo Andr\'e, São Paulo, Brazil}

\author{Stefano Sanvito\orcid{0000-0002-0291-715X}}\email{sanvitos@tcd.ie}
\affiliation{School of Physics and CRANN Institute, Trinity College Dublin, Dublin 2, Ireland}
\date{\today}

\begin{abstract}
Machine-learning models can be trained to predict the converged electron charge density of a density functional theory (DFT) calculation. In general, the value of the density at a given point in space is invariant under global translations and rotations having that point as a center. Hence, one can construct locally invariant machine-learning density predictors. However, the widely used projector augmented wave (PAW) implementation of DFT requires the evaluation of the one-center augmentation contributions, that are not rotationally invariant. Building on our recently proposed Jacobi-Legendre charge-density scheme, we construct a covariant Jacobi-Legendre model capable of predicting the local occupancies needed to compose the augmentation charge density. Our formalism is then applied to the prediction of the energy barrier for the 1H-to-1T phase transition of two-dimensional MoS$_2$. With extremely modest training, the model is capable of performing a non-self-consistent nudged elastic band calculation at virtually the same accuracy as a fully DFT-converged one, thus saving thousands of self-consistent DFT steps. Furthermore, at variance with machine-learning force fields, the charge density is here available for any nudged elastic band image, so that we can trace the evolution of the electronic structure across the phase transition. 
\end{abstract}

\maketitle

\section{INTRODUCTION}

The accurate prediction of material properties is the ultimate goal of computational 
materials science and one of the key enablers for new materials development. Density 
functional theory (DFT) \cite{Hohenberg1964,Kohn1965} is today the most widely used 
\textit{ab initio} method to compute materials properties. The DFT's success is due, 
among other reasons, to the very favourable trade-off between accuracy and computational
overhead~\cite{Burke_PerspectiveDFT2012}, a trade-off that can be fine-tuned to the 
system of interest by selecting the most appropriate exchange-correlation energy
functional~\cite{Perdew2001} and numerical implementation. 

The absence of an accurate density functional for the non-interacting kinetic energy 
restricts the direct minimization of the DFT energy to the solution of a set of 
single-particle equations, known as the Kohn-Sham (KS) equations \cite{Kohn1965}. 
These enter an iterative process, known as the self-consistent field (SCF) method, 
where at convergence the charge density defining 
the single-particle potential must equal that computed by solving the KS equations. 
Such self-consistent procedure constitutes the main numerical overhead of a DFT 
calculation. Any effort aiming at reducing the number of steps in a self-consistent 
cycle, or eliminating them completely, will enable a significant computational 
speed up and consequently will enhance the throughput. Machine learning (ML) may 
represent a possible avenue for reaching such a goal.

In recent years, ML has become an important tool in computational materials science, with 
applications distributed over a wide range of tasks~\cite{MLReview}. ML force 
fields~\cite{MLFFReview} are DFT-accurate energy models enabling large-scale task farming 
such as those needed in molecular dynamics simulations \cite{Caro2020,Smith2021,Willman2022}, 
crystal structure prediction \cite{Podryabinkin2019,Pickard2022} or convex-hull-diagram 
construction \cite{Gubaev2019,Wang2022,Minotakis2023,Roberts2024,Rossignol2024,Minotakis2023}. 
Alternatively, ML models can be constructed on either experimental or theoretical data, or on 
a combination of both, to enable rapid materials screening and the formulation of generative 
algorithms \cite{NelsonSanvito2019_MLCurieTemp,Focassio2021_MLPropPred_topo,Cobelli2022_MLMatDiscovery,Zipoli2022}. 
Most importantly for this discussion, ML has been used to augment the development of DFT
itself~\cite{Pederson2022}. The numerical construction of quantum-chemistry-accurate DFT 
functionals for specific materials 
\cite{Brockherde2017_burkeDatasetsBenzene,Nagai2020_MLXC,Li2021a_MLXC,DeepMind_MLXC_google} and 
for model Hamiltonian \cite{NelsonSanvito2019_MLDFTHubbard,Ryczko2019,Verdozzi2019,Moreno2020}, 
are just two examples. 

Importantly, as the Hohenberg-Kohn theorem~\cite{Hohenberg1964} establishes a one-to-one 
correspondence between the electron charge density and the external potential, it is clear 
that the knowledge of the atomic structure, determining the potential, should be sufficient 
to obtain the density. This means that, in principle, one can define a ML model that uses 
structural information to construct the DFT charge density. Such models can then be implemented 
for approximated exchange-correlation functionals, since the necessary training set can be 
generated by running standard DFT calculations. Several of such ML charge density models 
have been proposed, where either a local-orbital 
\cite{Grisafi2021_SAGPR_transferable_localenvironments,Lewis2021_SALTED,Zapeda2021,Rackers2023} 
or a real-space 
\cite{Chandrasekaran2019_ramprasadNNchg_ldos,Ellis2021_AttilaMALA,Lv2023,JLCDM_npjcomput_mat} 
representation of the charge density is used together with various ML algorithms. Once the 
converged ground-state charge density is generated, the associated observables (e.g. the 
dipole moment of a molecule) can be computed either directly~\cite{Sunshine2023,Grisafi2023} 
or through additional ML models using the density as an input~\cite{delRio2023}. A second
possibility is to use the ML-computed charge density as the initial density of a new Kohn-Sham 
self-consistent cycle. For an extremely accurate density, no further self-consistency will be 
needed and all the quantities available from KS-DFT will be computed without any further 
numerical effort, except for a single-shot solution of the Kohn-Sham equations. Otherwise, 
even for non-ultra-accurate ML models, one can still expect the ML density to be a convenient
starting point of a reduced self-consistent cycle. In both cases, the ML construction of the 
charge density can be integrated as an accelerator in any standard theoretical study involving 
DFT. 

Based on our Jacobi-Legendre (JL) cluster expansion for machine-learning force 
fields~\cite{Domina2023}, we have recently developed an efficient scheme to construct 
the DFT ground-state charge density over a real space grid~\cite{JLCDM_npjcomput_mat}. 
This predicts highly accurate charge densities, while demanding an extremely limited 
number of DFT calculations to perform the training. Most importantly, the model is 
constructed over an intrinsic many-body representation (the JL expansion), whose accuracy 
and complexity can be systematically tuned, but it is linear and thus lean. This means 
that the computational overheads for both training and inference remain very competitive 
and a parallel implementation is trivial. Although such JL charge density model can be
implemented with any DFT code writing the electron density over a real-space grid, 
it is currently implemented for the valence density obtained by the VASP
package~\cite{Kresse1996,Kresse1996c}. This uses the projector augmented wave (PAW) 
formalism, which is an efficient method to deal with the rapidly varying wavefunctions 
close to the atomic nuclei \cite{Blochl_1994,Kresse1999a}, and it is implemented in 
a multitude of DFT packages 
\cite{Kresse1996,Kresse1996c,Giannozzi_2009,Giannozzi_2017,Mortensen2005,Gonze2020}. 

In the PAW scheme, the total all electron wavefunction is written as a sum of two 
components, one that can be represented over a sparse Fourier/real-space grid and 
the other that is expanded over a dense atom-centered real-space grid close to the 
nuclei. These separated components are not independent from each other and are 
updated simultaneously during the SCF iterations. The separation in the wavefunction 
is also inherited by the charge density, so that both components are needed to 
construct the density-dependent Hamiltonian. 
In particular, in the PAW scheme it is convenient to work with the 
charge-compensated density, whose atom-centered component is completely determined 
by the PAW augmentation occupancies~\cite{Kresse1999a}. 
These are, therefore, essential to restart a VASP calculation. Hence, an ML model 
predicting the total density, then capable of being integrated with a PAW DFT workflow, 
should provide both the real-space charge-density component and the PAW occupancies.

This work generalizes our JL charge-density model~\cite{JLCDM_npjcomput_mat} to the 
prediction of the PAW occupancies. The main difference is that, while the charge density 
at a point in space is invariant for rotations of the local chemical environment about 
that point, the PAW occupancy is only covariant. Hence, here we first provide a general 
formulation of a covariant JL cluster expansion (Section \ref{MethodG}), and then we show 
how this can be used for predicting the PAW occupancies (Section \ref{MethodPAW}). The method 
is then applied to the calculation of the transition barrier between the 1H and 1T phases 
of MoS$_2$. After training over a very limited number of DFT calculations (Section \ref{Dataset}),
we will show that a combined JL charge-density and JL PAW occupancies model enable 
non-self-consistent nudged-elastic-band simulations at the same accuracy of fully converged 
ones, but at a tiny fraction of the computational costs (Section \ref{NEB}). Finally, we 
will provide some conclusions and an outlook on the potential of our scheme for materials 
design (Section \ref{Conclusion}). The paper is then complemented by three appendices, 
providing details of the DFT calculations, information about the hyperparameters optimization, 
and a pointer to our datasets.

\section{Methods}
\label{Method}

\subsection{Covariant Jacobi-Legendre cluster expansion}
\label{MethodG}

The formalism that we will develop here takes directly from our recently formulated 
Jacobi-Legendre potential (JLP) \cite{Domina2023} and Jacobi-Legendre charge-density 
model (JLCDM) \cite{JLCDM_npjcomput_mat}. 
In those two cases, the cluster expansion was constructed for quantities,
the atomic energy and the charge density at a grid point, invariant for rotations of the chemical 
environment with respect to the point of interest. Here such class of models is generalized to 
covariant quantities, and we call the new formalism covariant Jacobi-Legendre (CJL) cluster 
expansion. 
Consider a target quantity, $T_i$, associated with the $i$-th atom in the system. This 
can be generally written as a many-body expansion,
\begin{equation}\label{JLCE}
    T_i(\hat{\mathbf{r}}_{gi}) = T_i^{(\text{1B})}(\hat{\mathbf{r}}_{gi}) + T_i^{(\text{2B})}(\hat{\mathbf{r}}_{gi}) + T_i^{(\text{3B})}(\hat{\mathbf{r}}_{gi}) + \ldots\:,
\end{equation}
where the superscripts represent the body order of the expansion, while the index $i$ labels 
the atoms. Here, $\hat{\mathbf{r}}_{gi}$ is the versor along the direction connecting the
$i$-th atom and the point $g$ in Cartesian space. Thus, as with the JLCDM, $T_i$ is expanded 
over a Cartesian space mapped onto a real-space grid, but now the quantity of interest is localized 
at the atomic positions.

The various terms in the expansion can be explicitly written as
\begin{align}
T_i^{(\text{1B})}(\hat{\mathbf{r}}_{gi}) &= a_0^{Z_i} \:,\label{eqn:f1b}\\
T_i^{(\text{2B})}(\hat{\mathbf{r}}_{gi}) &= \sum_{j\neq i} \sum_{n l} a_{nl}^{Z_jZ_i}\overline{P}^{(\alpha,\beta)}_{nji}P_l^{gji}\:, \label{eqn:f2b}\\
T_i^{(\text{3B})}(\hat{\mathbf{r}}_{gi}) &= \sum_{(j,k)_i}\sum^\text{unique}_{\substack{n_1 n_2\\l_1 l_2 l_3}} a_{\substack{n_1 n_2\\l_1 l_2 l_3}}^{Z_kZ_jZ_i}\times\nonumber\\
& \times\sum_\text{symm}\left(\overline{P}^{(\alpha,\beta)}_{n_1ji}\overline{P}^{(\alpha,\beta)}_{n_2ki}P_{l_1}^{gji}P_{l_2}^{gki} P_{l_3}^{jki}\right)\:. \label{eqn:f3b}
\end{align}
In the equations \eqref{eqn:f1b}, \eqref{eqn:f2b} and \eqref{eqn:f3b}, we have used the vanishing Jacobi polynomials, $\widetilde{P}_{nji}^{(\alpha,\beta)}$, defined as 
\begin{equation}
\widetilde{P}_{nji}^{} = \left\{\begin{matrix*}[l]
P_n^{}\left(x_{ji}\right) - P_n^{}(-1) & \text{for } -1 \leq x_{ji} \leq 1 \\
0 & \text{for } x_{ji} < -1 \\
\end{matrix*}\right.
\label{eqn:shifted_jacobi_poly}
\end{equation}
and the double-vanishing-Jacobi polynomials, $\overline{P}_{nji}^{(\alpha,\beta)}$, defined as 
\begin{equation}
\overline{P}_{nji}^{} = \widetilde{P}_{n}^{}(x_{ji}) - \displaystyle\frac{\widetilde{P}_{n}^{}(1)}{\widetilde{P}_{1}^{}(1)}\widetilde{P}_{1}^{}(x_{ji}) \quad \text{for } n\geq 2 \:, \label{eqn:vanishing_jacobi_poly}
\end{equation}
where for simplicity of notation we have omitted the parameters $\alpha$ and $\beta$, which define 
the specific shape of the Jacobi polynomials of order $n$, $P_n^{(\alpha,\beta)}$. In the expression 
above we have also introduced $x_{ji}=\cos \left( \pi \dfrac{r_{ji} - r_{\rm min}}{r_{\rm cut} - r_{\rm min}} \right)$, 
with $r_{\rm min}$ and $r_{\rm cut}$ the minimum and cutoff radius, respectively, and the Legendre 
polynomial $P_l^{gji} = P_l(\hat{\mathbf{r}}_{gi} \cdot \hat{\mathbf{r}}_{ji})$. Note that up to the 
two-body order term, $T_i^{(\text{2B})}$, no symmetries must be explicitly included, since the central 
atom is distinct from the others. Then, for the three-body term, $T_i^{(\text{3B})}$, symmetries are 
relative only to the exchange of the atoms $j$ and $k$. 

In general, a given function can be written over its harmonics components as 
\begin{equation}
    f(\hat{\mathbf{r}}) = \sum_{lm} f_{lm}Y_{lm} (\hat{\vb r})\:, 
\end{equation}
where we have assumed that $f(\hat{\vb r})$ is real so that an expansion over real spherical 
harmonics, $Y_{lm} (\hat{\vb r})$, will return real radial components, $f_{lm}$. This assumption 
is not necessary and can be released by taking complex coefficients or complex spherical harmonics. 
The expansion coefficients are evaluated as,
\begin{equation}\label{Eq8}
    f_{lm} = \int f(\hat{\mathbf{r}}) Y_{lm} (\hat{\vb r}) \dd \hat{\vb r}\:,
\end{equation}
namely by projecting over the required angular momentum. Note that the integral in Eq.~(\ref{Eq8}) 
is over the solid angle spanned by the versor $\hat{\vb r}$. Let us now apply to our function of 
interest, $T_i(\hat{\vb r}_{gi})$, the same expansion, whose radial components can be computed 
by projection,
\begin{equation}
    T_{i,lm} = \int T_i(\hat {\vb r}_{gi}) Y_{lm} (\hat{\vb r}_{gi})\dd \hat{\vb r}_{gi} \:.
\end{equation}
The linearity of the expansion in Eq.~(\ref{JLCE}) establishes that the coefficients $T_{i,lm}$ 
can be written as the sum of different body-order contributions.

For the 1-body term, $T_{i,lm}^{(\text{1B})}(\hat{\mathbf{r}}_{gi})$, using Eq. \eqref{eqn:f1b}, the 
angular integrals all vanish, except that for $l=0$,
\begin{align}
T^{(\text{1B})}_{i,lm} &= \int T^{(\text{1B})}_i(\hat{\vb r}_{gi}) Y_{lm} (\hat{\vb r}_{gi})\dd \hat{\vb r}_{gi}=\nonumber\\
     &=\delta_{l0}\delta_{m0}\sqrt{4\pi} a_0^{Z_i} \:.
\end{align}
This corresponds to a contribution that arises only in the scalar (spherically symmetric) scenario.

In order to evaluate the 2-body term, $T_{i,lm}^{(\text{2B})}$, we consider the addition theorem for spherical harmonics (decomposition of a Legendre polynomial over spherical harmonics), which reads
\begin{equation}
    P_{l}^{gji} = \dfrac{4\pi}{2l+1}\sum_{m} Y_{l m}(\hat{\vb r}_{gi})Y_{l m}(\hat{\vb r}_{ji})\:.
\end{equation}
Then, by using the orthogonality condition,
\begin{equation}
        \int Y_{lm}(\hat{\vb r}_{gi})Y_{l'm'}(\hat{\vb r}_{gi})\dd \hat{\vb r}_{gi} = \delta_{ll'}\delta_{mm'}\:,
\end{equation}
we obtain
\begin{align}
T^{(\text{2B})}_{i,lm} &= \int T^{(\text{2B})}_{i}(\hat{\vb r}_{gi}) Y_{lm} (\hat{\vb r}_{gi})\dd \hat{\vb r}_{gi}=\nonumber\\
\qquad&=\dfrac{4\pi}{2l+1}\sum_{j\neq i} \sum_{n} a_{n l}^{Z_jZ_i}\overline{P}^{(\alpha,\beta)}_{nji}Y_{lm}( \hat{\vb r}_{ji})\:.
\end{align}
Note that this term is fully covariant, since, under rotations, it follows the same transformation 
rules of the spherical harmonics.

Finally, the 3-body term, $T_{i,lm}^{(\text{3B})}$, is evaluated in a similar way, by expanding the 
Legendre polynomials that depend on the grid point with the spherical-harmonics addition theorem. One 
can then perform the integration 
\begin{equation}
     \int Y_{l_1m_1}(\hat{\vb r}_{gi})Y_{l_2m_2}(\hat{\vb r}_{gi})Y_{lm}(\hat{\vb r}_{gi})\dd \hat{\vb r}_{gi} = \prescript{R}{}G^{l_1 l_2 l}_{m_1 m_2 m}\:,
\end{equation}
where we have used the real Gaunt symbols, $\prescript{R}{}G^{l_1 l_2 l}_{m_1 m_2 m}$. Thus, we obtain
\begin{align}
T^{(\text{3B})}_{i,lm} &= \int T^{(\text{3B})}_i(\hat{\vb r}_{gi}) Y_{lm} (\hat{\vb r}_{gi})\dd \hat{\vb r}_{gi}=\\
 &= \sum_{(j,k)_i}\sum^\text{unique}_{\substack{n_1 n_2\\l_1 l_2 l_3}} \dfrac{(4\pi)^2}{(2l_1+1)(2l_2+1)} a_{\substack{n_1 n_2\\l_1 l_2 l_3}}^{Z_jZ_kZ_i}\times\nonumber\\
 & \quad\times\sum_\text{symm}\bigg(\overline{P}^{(\alpha,\beta)}_{n_1ji}\overline{P}^{(\alpha,\beta)}_{n_2ki}P_{l_3}^{jki}\times \nonumber\\
 &\quad\times\sum_{m_1m_2}\prescript{R}{}G^{l_1 l_2 l}_{m_1 m_2 m}  Y_{l_1m_1} (\hat{\vb r}_{ji}) Y_{l_2m_2} (\hat{\vb r}_{ki}) \bigg)\:, \nonumber
\end{align}
where the second sum runs over the unique indexes $n_1$, $n_2$, taking care that the coefficient 
$a_{\substack{n_1 n_2\\l_1 l_2 l_3}}^{Z_jZ_kZ_i}$ is invariant under the simultaneous exchange of 
the species indexes $Z_j \leftrightarrow Z_k$ and the Jacobi indexes $n_1 \leftrightarrow n_2$, 
and therefore runs over non-equivalent coefficients ($n_1\geq n_2$). The third sum takes care of 
the symmetries ($\text{symm}$) by manually evaluating cases with equivalent coefficients with 
respect to $n_1\leftrightarrow n_2$. More details on the symmetries of the coefficients can be 
found in Ref. \cite{Domina2023}. We remark here that the pre-factor of the expansion coefficients, 
$a_{\substack{n_1 n_2\\l_1 l_2 l_3}}^{Z_jZ_kZ_i}$, is not affected by the symmetrization operation, 
since this involves the simultaneous exchange of the indexes $n_1\leftrightarrow n_2$ and 
$l_1\leftrightarrow l_2$, which leaves $T^{(\text{3B})}_{i,lm}$ unaffected.

This procedure creates a recipe to compute the harmonic components for any order of the expansion. 
The final result for the $lm$ harmonic component of the function $T_i$ is given by the sum of all 
the terms obtained, namely
\begin{equation}
    T_{i,lm} = T^{(\text{1B})}_{i,lm} + T^{(\text{2B})}_{i,lm} + T^{(\text{3B})}_{i,lm} + \ldots \label{eqn:harmonic_sum}
\end{equation}
where the 1-, 2- and 3-body components are now defined as,
\begin{align}
T^{(\text{1B})}_{i,lm} &= \delta_{l0}\delta_{m0}\sqrt{4\pi} a_0^{Z_i} \:,\label{eqn:t1}\\
T^{(\text{2B})}_{i,lm} &= \dfrac{4\pi}{2l+1} \sum_{j\neq i} \sum_{n} \bigg[ \delta_{l0} \frac{1}{\sqrt{4\pi}} a_{n0}^{Z_jZ_i}\widetilde{P}^{(\alpha,\beta)}_{nji} + \nonumber \\
&+ (1-\delta_{l0}) a_{nl}^{Z_jZ_i}\overline{P}^{(\alpha,\beta)}_{nji}Y_{lm}( \hat{\vb r}_{ji}) \bigg] \label{eqn:t2}\:,\\
T^{(\text{3B})}_{i,lm} &= \delta_{l0} \sum_{(j,k)_i}\sum^\text{unique}_{n_1 n_2 l_1} (4\pi)^2 a_{n_1 n_2l_1}^{Z_jZ_kZ_i} \times \nonumber \\
& \times \sum_\text{symm}\bigg(\overline{P}^{(\alpha,\beta)}_{n_1ji}\overline{P}^{(\alpha,\beta)}_{n_2ki}P_{l_1}^{jki}\bigg) + \nonumber\\
&+(1-\delta_{l0}) \sum_{(j,k)_i}\sum^\text{unique}_{\substack{n_1 n_2\\l_1 l_2 l_3}} \dfrac{(4\pi)^2}{(2l_1+1)(2l_2+1)} a_{\substack{n_1 n_2\\l_1 l_2 l_3}}^{Z_jZ_kZ_i}\times\nonumber\\
 & \quad\times\sum_\text{symm}\bigg(\overline{P}^{(\alpha,\beta)}_{n_1ji}\overline{P}^{(\alpha,\beta)}_{n_2ki}P_{l_3}^{jki}\times \nonumber\\
 & \quad\times\sum_{m_1m_2}\prescript{R}{}G^{l_1 l_2 l}_{m_1 m_2 m}  Y_{l_1m_1} (\hat{\vb r}_{ji}) Y_{l_2m_2} (\hat{\vb r}_{ki}) \bigg)\:.
 \label{eqn:t3}
\end{align}
One should notice that we separate the $l=0$ and $l>0$ contributions. Due to the discontinuity 
of the spherical harmonics at the origin, in $T^{(\text{2B})}_{i,lm}$, for $l=0$, we use the 
vanishing Jacobi polynomials, $\widetilde{P}^{(\alpha,\beta)}_{nji}$, defined in 
Eq. \eqref{eqn:shifted_jacobi_poly}, while for $l>0$, we use the double-vanishing Jacobi 
polynomials, $\overline{P}^{(\alpha,\beta)}_{nji}$, defined in Eq. \eqref{eqn:vanishing_jacobi_poly}, 
since these vanish at the origin. In doing so we keep the expansion continuous. 
In addition, actual implementations of this formalism can redefine the expansion coefficients so 
to absorb the pre-factors of Eqs. \eqref{eqn:t1}, \eqref{eqn:t2} and \eqref{eqn:t3}.

\subsection{Application to PAW augmentation charges}
\label{MethodPAW}

\begin{figure*}
\centering
\includegraphics[width=\linewidth]{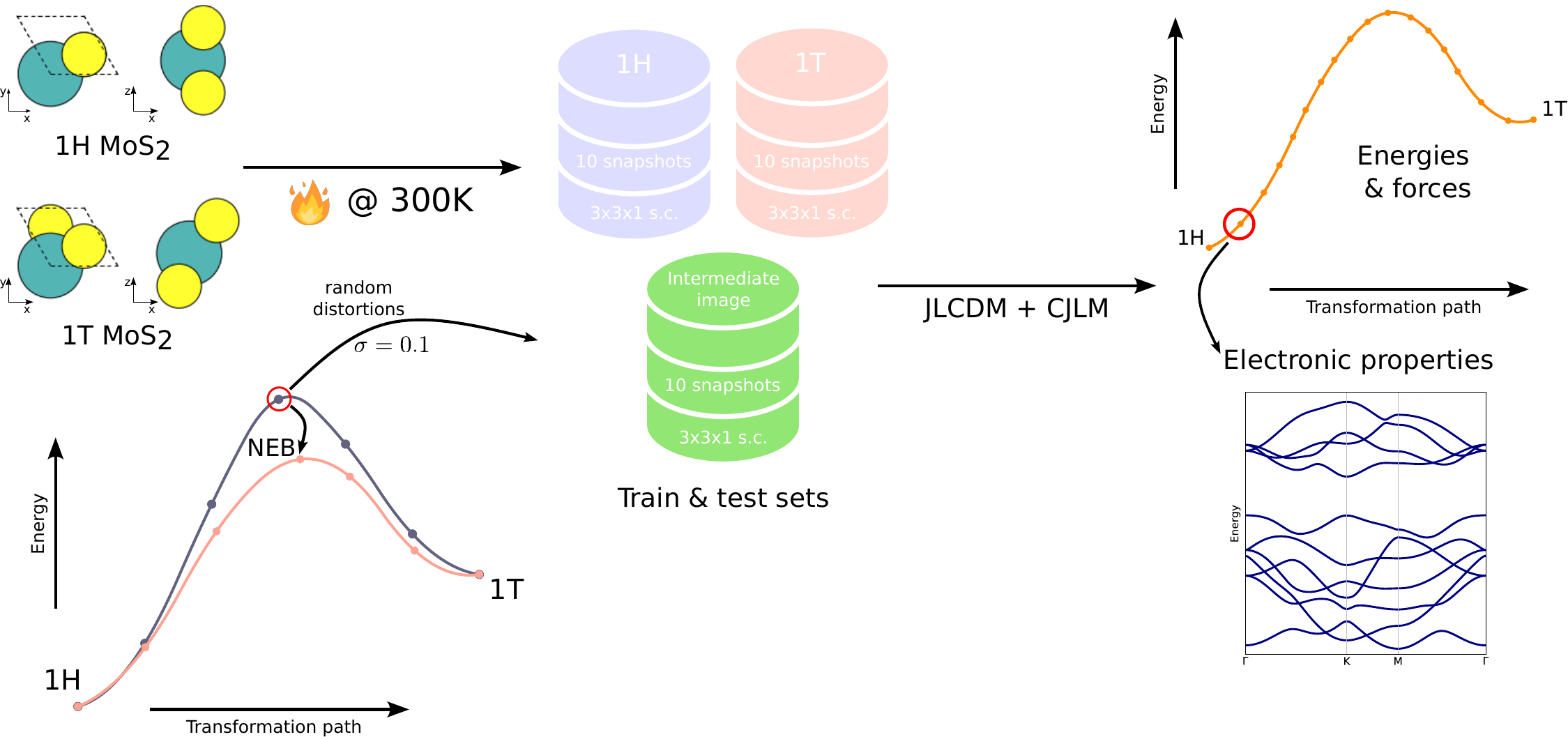}
\caption{Illustration of the workflow used to construct a JLCDM and CJL model to predict the 
converged DFT ground-state charge density and PAW occupancies. Both these ingredients are necessary 
to start a DFT PAW calculation (with VASP in this case), and here the method is used to predict the 
energy barrier for the 1H-to-1T structural transition of MoS$_2$. The creation of the dataset starts 
by computing the 1H and 1T phase of 2D MoS$_2$. We perform AIMD  at 300~K for each phase and extract 
10 snapshots of each, where we take a 1:1 ratio between training and test snapshots. Ten additional 
data points are created by using random distortions of linearly interpolated structures along the NEB 
path (before any optimization). This training set is used to construct the JLCD and CJL models, which 
are then used to predict the charge density and PAW occupancies for any distorted structure with the 
MoS$_2$ chemistry. The models can be readily used to predict the entire NEB trajectory and to investigate 
properties such as the electronic structure along the transition path.}
\label{fig:workflow}
\end{figure*}

Our goal is to reconstruct the augmentation (compensation) charge density, $\widehat{n}$, which is 
defined as
\begin{equation}
    \widehat{n}(\mathbf{r}) = \sum_{ij,LM} \rho_{ij}\widehat{Q}_{ij}^{LM}\:, \label{eqn:compensation_density2}
\end{equation}
where $\rho_{ij}$ is the occupancy of the augmentation channel $(i,j)$ and $\widehat{Q}_{ij}^{LM}$ 
is the $L$-dependent compensation charge. 
We use here the same notation as that in Ref.~\cite{Kresse1999a}, to distinguish the charge density 
$n(\mathbf{r})$ and the augmentation charge density $\widehat{n}(\mathbf{r})$. Following the same 
convention, the indexes $i$ and $j$ represent the collection of labels that are required to fully 
characterize an augmentation channel, as shown below, and in this context they must not be confused 
with labels for atomic positions. In order to cluster expand the compensation charge of Eq.~\eqref{eqn:compensation_density2}, it is useful to write it with respect to the site $\mu$ at 
position $\mathbf{R}_{\mu}$,
\begin{equation}
    \widehat{n}_{\mu}(\mathbf{r}) = \sum_{ij,LM} \rho_{ij} q_{ij}^{LM}\,g_{L}(|\mathbf{r}-\mathbf{R}_{\mu}|)Y_{LM}(\widehat{\mathbf{r}-\mathbf{R}_{\mu}})\:,
\end{equation}
where we have used the expansion, 
\begin{equation}
    \widehat{Q}_{ij}^{LM} = q_{ij}^{LM}\,g_{L}(|\mathbf{r}-\mathbf{R}_{\mu}|)Y_{LM}(\widehat{\mathbf{r}-\mathbf{R}_{\mu}})\;,
\end{equation}
\noindent with $g_L$ be a linear combination of two spherical Bessel functions \cite{Kresse1999a}, 
while $Y_{LM}$ be spherical harmonics. Here the $q_{ij}^{LM}$ terms act as expansion coefficients 
of the compensation charge $\widehat{Q}^{LM}_{ij}$, with respect these functions. By writing explicitly 
the indexes $i$ and $j$ in terms of $(k_1,l_1,m_1)$ and $(k_2,l_2,m_2)$, respectively, we obtain 
\begin{equation}
    \widehat{n}_{\mu}(\mathbf{r}) = \sum_{\substack{{k_1 k_2}\\{LM}}} \widehat{n}_{\mu k_1 k_2}^{LM}\,g_{L}(|\mathbf{r}-\mathbf{R}_{\mu}|)Y_{LM}(\widehat{\mathbf{r}-\mathbf{R}_{\mu}})\:,
\end{equation}
where we have defined 
\begin{equation}
    \widehat{n}_{\mu k_1 k_2}^{LM} = \sum_{\substack{{l_1 l_2}\\{m_1 m_2}}} \rho_{\substack{{k_1 l_1 m_1}\\{k_2 l_2 m_2}}} q_{\substack{{k_1 l_1 m_1}\\{k_2 l_2 m_2}}}^{LM}\:.
\end{equation}
These are the harmonics components of $\widehat{n}_\mu$, belonging to the subspace of angular 
momentum $(L,M)$. 
The specific nature of each term in the product is unessential for the discussion 
here, and for further details we refer to the PAW formulation contained in reference \cite{Kresse1999a}. 
In particular, for each choice of pairs $(k_1, k_2)$, we can design covariant tensor models that fits the 
$(L,M)$ harmonic components of $\widehat{n}_\mu$, namely
\begin{equation}
     T_{\mu,LM} \simeq \sum_{\substack{{l_1 l_2}\\{m_1 m_2}}} \rho_{\substack{{k_1 l_1 m_1}\\{k_2 l_2 m_2}}} q_{\substack{{k_1 l_1 m_1}\\{k_2 l_2 m_2}}}^{LM}\:.
\end{equation}

The general case would require fitting an expression for 
\begin{equation}
    \widehat{n}_{\mu k_1 k_2}^{LM} = T_{\mu,LM}^{(\text{1B})} + T_{\mu,LM}^{(\text{2B})} + T_{\mu,LM}^{(\text{3B})} + \ldots,
\end{equation} 
however, we resort to a simplification to reduce the total number coefficients involved in the
expansion. Our strategy now is to distinguish three cases. The first case is $l_1 = l_2 = 0$, 
resulting in $L = M = 0$. For this the appropriate tensor components can be derived directly from 
Eqs. \eqref{eqn:t1} and \eqref{eqn:t2}, restricting this expansion to 1 and 2-body terms,
\begin{equation}
\widehat{n}_{\mu 0 0}^{0 0} = a_0^{Z_i} + \sum_{j\neq i} \sum_{n} a_{n}^{Z_jZ_i}\widetilde{P}^{(\alpha,\beta)}_{nji} \label{eqn:n_00_00}\:,
\end{equation}
where we have incorporated the constant leading factor $\sqrt{4\pi}$ into the $a$'s coefficients 
to be fitted. This case is similar to fitting a force field, namely, the component of the augmentation 
density has spherical symmetry.

Next, we consider the situation where either $l_1$ or $l_2$ are zero, but not both. For $l_2 = 0$, 
we have $l_1 = L \neq 0$. In this case, the quantity of interest is a tensor of order $L$. Therefore 
we can write the expression for $\widehat{n}_{\mu L0}^{LM}$  using only the 1-body term as
using Eq. \eqref{eqn:t2}
\begin{equation}
    \widehat{n}_{\mu L0}^{LM} = \frac{4\pi}{2L+1} \sum_{j\neq \mu} \sum_{n} \bigg[ a_{nL}^{Z_jZ_\mu}\overline{P}^{(\alpha,\beta)}_{nj\mu}Y_{LM}( \hat{\vb r}_{j\mu}) \bigg]\:. \label{eqn:n_l0_LM}\
\end{equation}

Finally, we have the most general case where $l_1 \neq l_2 \neq 0$. We can, then, write the harmonic 
components of the density using $T_{\mu,lm}^{(\text{3B})}$, but without performing the sum over $l_1$ 
and $l_2$ in order to properly account for the angular momentum coupling, namely for $T_{\mu,lm}^{(\text{3B})}$ 
to transform as $n_{\mu l_1,l_2}^{LM}$. Explicitly we have
\begin{align}
    n_{\mu l_1,l_2}^{LM} &= \sum_{(j,k)_\mu}\sum^\text{unique}_{\substack{n_1 n_2\\l^\prime}} \dfrac{(4\pi)^2}{(2l_1+1)(2l_2+1)} a_{\substack{n_1 n_2\\l_1 l_2 l^\prime}}^{Z_jZ_kZ_\mu}\times\nonumber\\
 & \times\sum_\text{symm}\bigg(\overline{P}^{(\alpha,\beta)}_{n_1j\mu}\overline{P}^{(\alpha,\beta)}_{n_2k\mu}P_{l^\prime}^{jk\mu}\times \nonumber\\
 & \times\sum_{m_1m_2}\prescript{R}{}G^{l_1 l_2 L}_{m_1 m_2 M}  Y_{l_1m_1} (\hat{\vb r}_{j\mu}) Y_{l_2m_2} (\hat{\vb r}_{k\mu}) \bigg)\:. \label{eqn:n_l1l2_LM}
\end{align}

The expressions derived in Eqs.~\eqref{eqn:n_00_00} through \eqref{eqn:n_l1l2_LM} can now be 
used to efficiently predict the PAW occupancies. This means that now the training of a JLCDM 
and a CJL model for the PAW occupancies gives us access to a full machine-learning-predicted 
charge density. Such density can then be used as a starting point for VASP calculations, and 
possibly as converged charge density for non-self-consistent DFT. 

\section{Results and Discussion}
\label{Results}

\begin{figure*}[hbt]
\centering
\includegraphics[width=\linewidth]{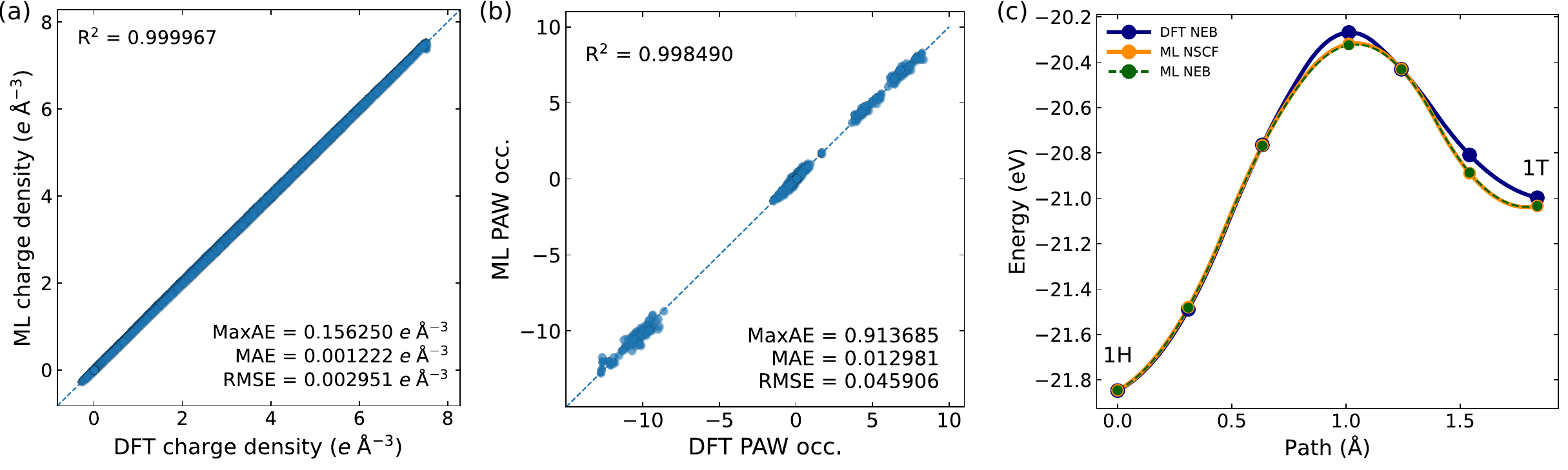}
\caption{Parity plots for the various density models constructed for MoS$_2$, namely (a) the 
JLCDM and (b) the CJL model. Data are for the test set and the values of the mean absolute error 
(MAE), root-mean-squared error (RMSE), maximum absolute error (MaxAE), and R$^2$ metrics are 
reported. Panel (c) displays the total energy along the transition path between the 1H and 1T 
phases. The total energy is obtained with NEB calculations using self-consistent DFT total energies 
and forces (DFT NEB), the ML models at the self-consistent DFT structures (ML NSCF), a fully 
ML-driven NEB, where the structures are relaxed with non-self-consistent DFT using the ML models 
(ML NEB). The solid line is a spline interpolation of the actual data (symbols). }
\label{fig:results_parity}
\end{figure*}

We now demonstrate the application of the JLCDM together with the CJL model into a typical 
materials science workflow. In particular, we study the structural transition between the 
1H and the 1T phase of 2D MoS$_2$ monolayer. We will use our machine-learning models to find 
the transition state and all the desired images along the transition without any self-consistent 
calculation, but rather, by evaluating non-self-consistently energies and forces. Furthermore, 
as here energy and forces are obtained from the knowledge of the full electron density, any 
electronic quantity is also readily available. This is, for instance, the case of the band 
structure along the transition. Our workflow is illustrated in Fig.~\ref{fig:workflow}.

\subsection{Dataset generation and ML models training}
\label{Dataset}

Firstly, as in any machine-learning pipeline, we need to generate the data to train the models. 
The training/test data consists of structures derived from three distinct starting configurations, 
namely the  1H phase, the 1T phase, and the geometrical midpoint interpolating the atomic Cartesian 
coordinates of the two phases. For the 1H and 1T phases, we take snapshots from \textit{ab initio} 
molecular dynamics (AIMD) simulations at 300~K. In contrast, the snapshots of the geometrical midpoint 
are obtained from random distortions (the standard deviation of the distortion amplitude is $0.1\;\text{\AA}$). 
For each category, we sample 10 distinct geometries and equally split them into training 
and test data (1:1 ratio), see Fig.~\ref{fig:workflow}. The grid point sampling of the valence charge 
density, described by the JLCDM, is performed with the same procedure established in 
Ref.~\cite{JLCDM_npjcomput_mat}, a pipeline demonstrated efficient. Then, the hyperparameter optimization 
of the JLCDM is carried out through Bayesian optimization using Gaussian Processes over the training set, 
see Table \ref{tab:model_hyperparameters} in Appendix B for more details. In contrast, when fitting the 
PAW components, we consider the following $l$-channels for Mo: $0,0,1,1,2,2$, and for S: $0,0,1,1$. This 
choice maps all the available components for each atom, following the rule $L= |l_i-l_j|,|l_i-l_j|+2,\ldots,l_i+l_j $, 
and generates 138 components for the Mo ions and 33 components for the S ions in each snapshot.

In general, the transition state (TS) is computed with the nudged elastic band (NEB) 
method \cite{neb1,neb_tangent} with climb image \cite{neb_climb_image} as implemented in the ASE package
\cite{ase_HjorthLarsen_2017}. The energy and the atomic forces needed by the NEB are computed with the 
VASP code \cite{Kresse1996, Kresse1996c}. Conventionally, all DFT calculations are performed self-consistently, 
but here we will replace those with a non-self-consistent evaluation using our predicted densities. In any 
case, the starting point of any NEB calculation is the linear interpolation of the initial (1H-MoS$_2$) 
and final (1T-MoS$_2$) structure into intermediate structures, also referred to as images. For our specific 
problem, we construct 5 images across the transition.

In closing this section we wish to spend a few words on our strategy to generate the models
training set. Currently, this is pretty simple, namely we have included a few configurations around the two stable
phases and at the geometrical midpoint between them. The question is whether better choices are possible. 
In general, charge-density models are more subtle to train than machine-learning potentials, since a single DFT 
calculation generates a large number of training points, in principle all the grid points, but has also significant 
redundancy. Thus, significant data pruning must be implemented. Then, it is important to note that the final models 
are typically rather slim, contain a few thousands of parameters. This means, that simply expanding the number 
of training configurations, thus expanding the pool of training points, does not guarantee achieving better models. 
As a consequence, one has to design a strategy to include in the training set the configurations most representative 
for the problem at hand, effectively an active-learning strategy, here specific of a NEB calculation.

\subsection{Transition-state, energy, and electronic properties}
\label{NEB}

Figure \ref{fig:results_parity} presents the results for the fit of both the JLCDM for the density 
written over the real-space grid [panel (a)] and the CJL model for the PAW occupancies [panel (b)]. 
In general, we obtain an extremely accurate JLCDM, with a root-mean-squared error (RMSE) of only 
$2.9\cdot 10^{-3}$~$e$/{\AA}$^3$, a mean absolute error (MAE) of $1.2\cdot 10^{-3}$~$e$/{\AA}$^3$ 
and a maximum absolute error (MaxAE) of $0.15$~$e$/{\AA}$^3$. These are a factor 2 to 3 lower than 
those of our previously published JLCDM for the same compound~\cite{JLCDM_npjcomput_mat}. Such improvement 
is related to the slightly expanded training set and to the nature of the test set. In fact, we now 
train also on interpolated structures along the 1H-1T transition, and not just over the 1H and 1T phases, 
and we test on different images across the phase transition (in Ref.~\cite{JLCDM_npjcomput_mat} the 
test set was formed by distorted images of the 1T$^\prime$ structure).

The same parity plot for the CJL model is presented in Fig.~\ref{fig:results_parity}(b), again for 
the test set. Here we aggregate all the PAW components, that is all the $(l_i,l_j,L,M)$ components 
for all Mo and S atoms. Also in this case the model, which interpolates with only 143 parameters 
(see Appendix B), returns extremely accurate predictions, with a RMSE of 0.045, a MAE of 0.012 and 
a MaxAE of 0.91. Thus, the two models combined produce an extremely accurate total charge density 
that can be readily used to evaluate energy and forces. 

Such evaluation is provided in Fig.~\ref{fig:results_parity}(c), where we present the potential 
energy surface (PES) across the 1H to 1T transition. In particular, the calculation is performed 
in three different ways. The ground truth is provided by a fully self-consistent NEB calculation 
(`DFT NEB', blue line), where each of the images along the path is fully relaxed by using self-consistent 
DFT with the standard charge-density initialization. Typically, this requires the evaluation of about 1,000 
structures in total, so that the computational overheads are that of 1,000 self-consistent DFT calculations. 
The first test for our ML models is performed over the total energy. For this we take the NEB transition 
states computed with self-consistent DFT and evaluate their total energy by using the ML density and no 
self-consistent iteration (`ML NSCF', yellow line). Next, we assess both energy and forces, namely we 
use the ML models to perform the entire NEB workflow (`ML NEB', green line). In this case, we input the 
JLCDM-predicted charge density and CJL-predicted PAW occupations for each structure at each step of 
the NEB optimization. For these, energy and forces are calculated with non-self-consistent DFT. Thus, 
in this last part, both the relaxation and the energy evaluation are driven by our charge-density models. 
This second test is clearly more stringent, since errors may be present in both the energy and the 
structure evaluation and they can add up. However, should the predictions be valid, one will be able 
to perform an entirely non-self-consistent PES evaluation, thus saving the computational cost of all 
the self-consistent iterations involved in the NEB workflow.

In general, we find an excellent agreement between both our ML-driven PES evaluations and the ground 
truth, with minimal differences associated to the structural evaluation. This means that the ML-driven 
NEB relaxes at practically the same structures obtained with fully self-consistent DFT, and the remaining 
error is attributed to the energy of the final structures. More in detail, the models seem to perform 
better at the 1H side of the PES than at the 1T one. The height of the energy barrier, namely the energy 
difference between the lowest energy phase (1H) and the transition state, is computed at 1.5270~eV 
and 1.5212~eV, respectively for ML NSCF and ML NEB, against a self-consistent DFT energy of 1.5783~eV.
Thus, the error over the barrier height is 0.0512~eV for ML NSCF and 0.0570~eV for ML NEB. These are 
errors in the 50~meV range, corresponding to $\sim$3\% of the target self-consistent energy-barrier height. 
The same comparison for the difference in energy between the 1T phase and the transition state energy, 
returns us 0.7143~eV, 0.7085~eV, and 0.7286~eV, respectively for ML NSCF, ML NEB, and DFT NEB, with a 
deviation of 0.0142~eV for ML NSCF and of 0.020~eV for ML NEB. These are in the 20~meV range or about 2\%. 
Finally, the difference in energy between the 1H and 1T phase is almost identical for ML NSCF and ML NEB, 
0.8127~eV, and again it is extremely close to that computed by self-consistent DFT, 0.8497~eV. See 
Table~\ref{tab:results_energies} for a summary of the results. With these results at hand, we can conclude 
that the ML-computed density is certainly sufficient to replace the self-consistent cycle across the entire 
PES evaluation. 

\begin{table}[h]
\caption{Energies for the transition between the 1H and 1T phase of 2D MoS$_2$. DFT NEB is the result 
obtained by the full \textit{ab-initio} NEB calculation. ML NSCF refers to the case, where we compute 
the energy using ML non-self-consistent DFT at the DFT-relaxed structures. Finally, ML NEB refers to 
a completely ML-driven NEB (no self-consistent cycles are performed at any point). Here $\Delta E_{\alpha-TS}$ 
is the absolute energy difference between the $\alpha$ phase and the transition state, while 
$\Delta E_{1H-1T}$ is the absolute energy difference between the two phases.}
\label{tab:results_energies}
\begin{tabular}{lccl}
\hline
        & $\Delta E_{1H-1T}$ (eV) & $\Delta E_{1H-TS}$ (eV) & $\Delta E_{1T-TS}$ (eV) \\ \hline
DFT NEB & 0.8497 & 1.5783 & 0.7286 \\
ML NSCF & 0.8127 & 1.5270 & 0.7143 \\
ML NEB  & 0.8127 & 1.5212 & 0.7085 \\ \hline
\end{tabular}
\end{table}

\begin{figure*}
\centering
\includegraphics[width=\linewidth]{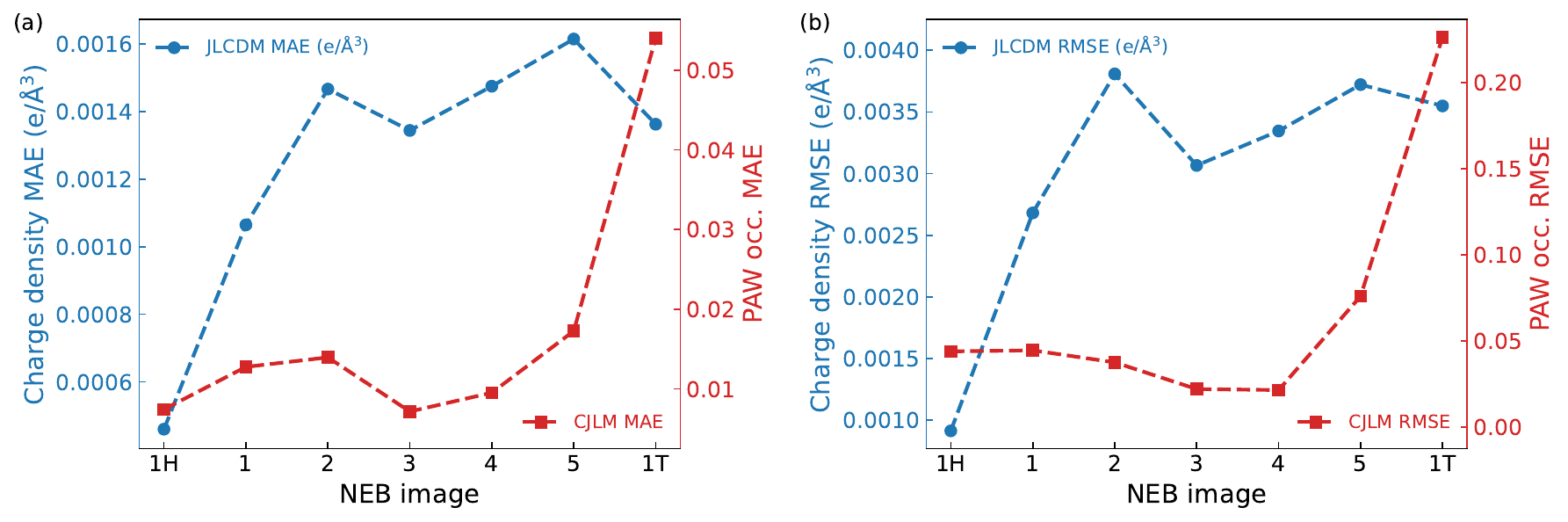}
\caption{(a) MAE and (b) RMSE for the two charge-density models computed for each of the NEB
images. Note that both density models perform better on the 1H side of the phase transition. However, while the
JLCDM away from the 1H phase has similar accuracy for all images, the CJL model has a sharp increase in the 
error for phases close to the 1T only.}
\label{fig:error}
\end{figure*}

Although small, it is interesting to understand the origin of the deviations between
the ML-computed NEB energy and that obtained by self-consistent DFT. This is certainly not attributable 
to differences in atomic structure that might have resulted from the different relaxation processes. In fact,
the energies computed with the ML density at the self-consistent DFT geometries are practically identical
to those computed along the ML-NEB curve. Then, the error has to be found in the charge density alone.
In order to understand better this point, in Fig.~\ref{fig:error} we plot the MAE and RMSE of both 
charge-density models for all the different NEB images. We find that, the JLCDM performs significantly
better for the 1H structure, but then displays a reasonably uniform error across the entire NEB path.
For instance, with the exception of the 1H structure the RMSE ranges between $2.6\cdot10^{-3}$~$e$/\AA$^3$ 
and $3.7\cdot10^{-3}$~$e$/\AA$^3$, while the MAE goes from $1.0\cdot10^{-3}$~$e$/\AA$^3$ to
$1.6\cdot10^{-3}$~$e$/\AA$^3$ (for the 1H structure the RMSE and MAE are $0.9\cdot10^{-3}$~$e$/\AA$^3$
and $0.4\cdot10^{-3}$~$e$/\AA$^3$, respectively.) In contrast, the CJL model for the PAW occupancies 
has a rather constant error across the NEB trajectory, except for the 1T phase and its closest image, where 
it spikes up. The error increase, in this case, is really significant with both RMSE and MAE of the 1T image 
being about a factor five larger than their average over the entire structural transition. 

The error distribution across the NEB path reflects in the energy error that, as noticed 
before, is larger on the 1T side of the transition. It is thus clear that, although the models constructed appear 
already accurate enough to enable an accurate evaluation of the transition barrier, a higher degree of accuracy 
can be reached by a more balanced training set. In particular, given the fact that the JLCDM and the CJL model
are rather different in size, different training sets for the two models may be appropriate.

\begin{figure*}
\centering
\includegraphics[width=\linewidth]{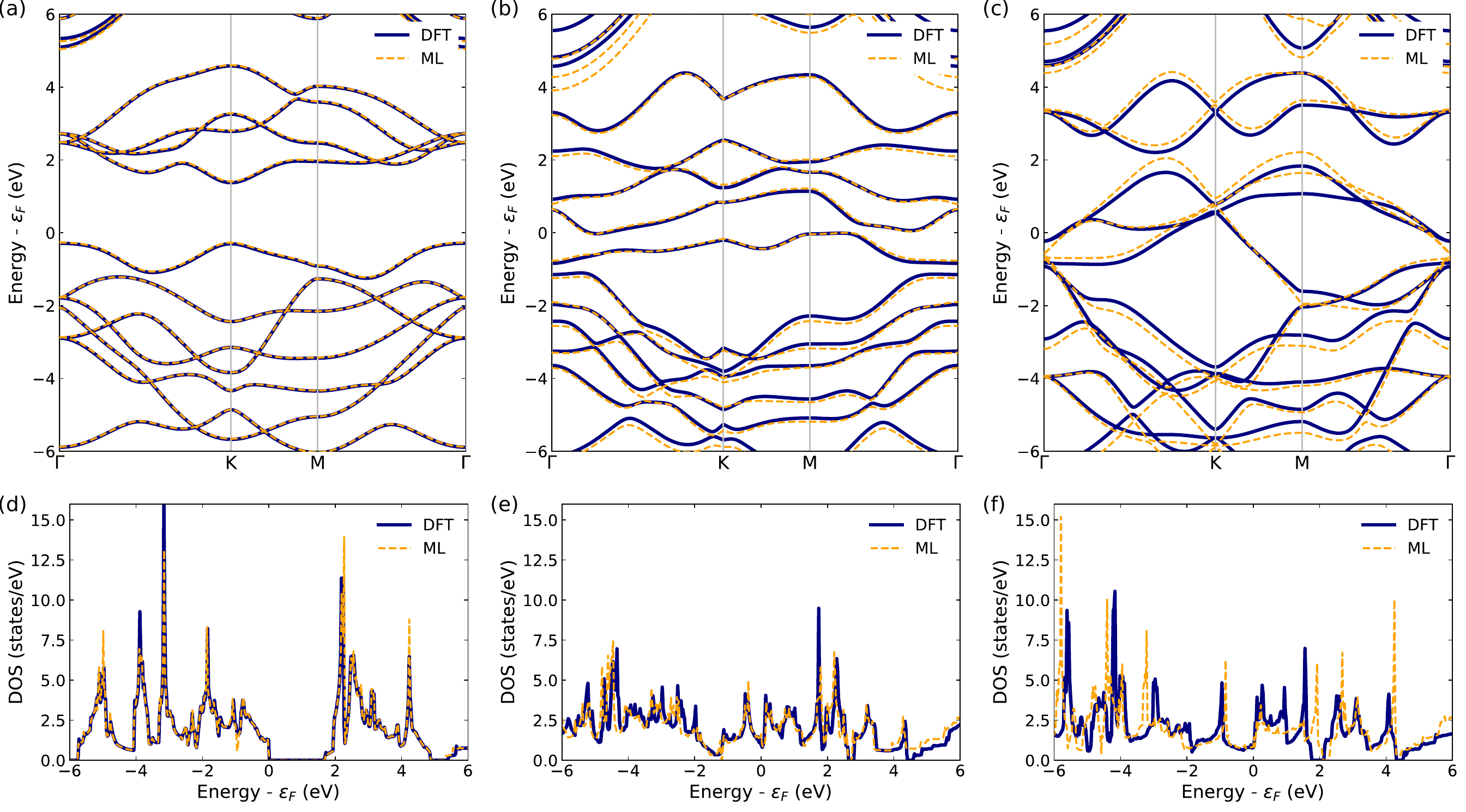}
\caption{Band structure and density of states (DOS) for MoS$_2$ along the 1H-to-1T transition. We 
compare results obtained from fully self-consistent DFT (black lines) and non-self-consistent DFT 
with charge density and PAW occupancies obtained from the JL models. Panels (a) and (d) are for 1H, 
(b) and (e) for the transition state, and (c) and (f) for the 1T phase.}
\label{fig:results_bands_dos}
\end{figure*}

At variance with ML force fields, the use of the charge density allows one, not only to compute 
energy and forces but also any other observable related to the charge density. This is, for instance, 
the case of the Kohn-Sham spectrum. In other words, we are here able to track the electronic structure 
of MoS$_2$ along the 1H-to-1T structural transition. Our results are presented in 
Figure~\ref{fig:results_bands_dos}, where we report the band structure and the DOS of the 1H phase, the 
1T one, and the structure corresponding to the transition state. In particular, we compare the fully 
self-consistent DFT electronic structure, with that obtained from the density predicted by our two 
charge-density models. In general, we find a rather good agreement between the DFT and the ML bands/DOS, 
an agreement that is certainly much more pronounced at the 1H side of the structural transition. This 
is a consequence of the better ML-interpolated charge density found for the 1H phase, as already observed 
in the discussion of the total energy. In any case, the ML models have enough accuracy to predict the 
insulator-to-metal transition along the structural transformation, with the transition state being 
already metallic. Interestingly, we find that the accuracy of the ML models is higher in predicting 
occupied states, with the largest deviation found for bands several eV away from the Fermi level. 
We also note that the bands computed with the ML density are more accurate at
the 1H side of the phase transition, as expected by the error distribution discussed before.

In any case, we can conclude that the ML charge density provides an excellent tool to approximate 
the electronic structure across structural deformations of a solid/molecule. Here we have considered 
a rather broad structural ensemble, since it spans across a phase transformation. Should one have a 
tighter pool of structures, as those sampled in a molecular dynamics simulation at moderate temperature, 
the agreement will be significantly higher. This will allow us to compute electronic-structure 
observables (e.g. the band gap or the optical absorption spectrum) at the DFT level, but at the 
computational cost of non-self-consistent DFT. We believe that such an advantage can open the possibility 
for an inexpensive evaluation of temperature-dependent materials properties from DFT.

\section{Conclusion}
\label{Conclusion}

Extending on our previously introduced JLCDM~\cite{JLCDM_npjcomput_mat} and JL force 
field~\cite{Domina2023}, we have here introduced the formalism for an atom-centered 
covariant Jacobi-Legendre cluster expansion. This is then applied to the prediction of 
the PAW occupations needed by a PAW-based DFT calculation to represent the full charge 
density. As such, the covariant cluster expansion together with the JLCDM provides a 
ML avenue to represent the entire DFT-PAW charge density, that can be used to prepare 
the starting density of a DFT calculation.

We have implemented such a scheme in the VASP code and tested it for 2D monolayer MoS$_2$. 
In particular, we have taken the challenging task of tracking the structural and electronic 
changes across a transition from the 1H to the 1T phases. This is a rather stringent test, 
with a significant amount of diversity in the structures to predict. We have found that, with 
a small training set (only 15 DFT calculations), our ML models were able to reproduce an entire 
NEB search and to track the electronic structure across the structural transformation. 
This allows us to save several fully converged self-consistent DFT calculations. 
Our results make charge-density models an attractive alternative to machine-learning force 
fields, since in addition to energy and forces they can reproduce the electronic structure 
over an ensemble of configurations. Applications related to temperature-dependent properties 
of materials are thus envisioned.

\begin{acknowledgments}
This work was supported by S\~ao Paulo Research Foundation (FAPESP) (Grants no. 2021/12204-6, 
2019/04527-0, and 2017/02317-2), and the Irish Research Council Advanced Laureate Award 
(IRCLA/2019/127). UP thanks additional funding from the Qatar National Research Fund (NPRP12S-0209-190063) 
and Science Foundation Ireland (19/EPSRC/3605).  We acknowledge the DJEI/DES/SFI/HEA Irish Centre 
for High-End Computing (ICHEC) and Trinity Centre for High Performance Computing (TCHPC) for the 
provision of computational resources. We acknowledge NVIDIA Academic Hardware Grant Program for 
providing graphics processing units. 
\end{acknowledgments}

\section*{Appendix A: DFT calculations}

All single-point and AIMD calculations are performed using DFT \cite{Hohenberg1964, Kohn1965} 
as implemented in the Vienna \textit{ab initio} simulation package (VASP) \cite{Kresse1996, Kresse1996c}. 
The exchange and correlation energy is provided by the generalized gradient approximation 
(GGA) \cite{Perdew1992} within the Perdew-Burke-Ernzerhof (PBE) \cite{Perdew1996a} formulation 
and parameterization. As already discussed, we use projector augmented wave (PAW) \cite{Kresse1999a} 
pseudopotentials. Single-point self-consistent calculations are performed with a \SI{600}{\electronvolt} 
kinetic-energy cutoff for the plane-wave expansion, and the Brillouin zone is sampled over a $k$-point 
density of $12\;/$\si{\per\angstrom}. 
AIMD calculations are performed with a \SI{2}{\femto\second} time-step, and the Nosé-Hoover 
thermostat \cite{Nose1984, Nose1984b, Hoover1985} at $300\;\rm K$. All AIMD runs are at 
least \SI{4}{\pico\second} long, and snapshots are taken from the simulation's last 
\SI{3}{\pico\second}. Random distortions applied to the atomic positions are sampled from 
a probability distribution using $\sigma=0.1$~\AA\ as the standard deviation.

Nudged elastic band (NEB) \cite{neb1,neb_tangent} calculations are performed with climb image 
\cite{neb_climb_image}, as implemented in the ASE package \cite{ase_HjorthLarsen_2017}. Energies 
and forces are calculated using the VASP code, either self-consistently or with the ML charge 
density and PAW occupations.

From the practical point of view our machine-learning charge density is handled externally 
from VASP, which has not been modified. In fact, we simply use the tools available in the Pymatgen~\cite{Pymatgen}
library to read/write the VASP CHGCAR file. This contains the lattice information and atomic positions, the charge
density component on the grid and the augmentation charges. Then, the JLCDM is used to predict the
charge density on the grid and the CJLM to compute the augmentation charges. These are written back
on the CHGCAR file by Pymatgen with the correct format.

\section*{Appendix B: Model training and hyperparameters}

The models are fitted by using singular value decomposition to find the pseudo-inverse of the 
matrix $A$ defining the equation $A\hat{x}=\hat{b}$ for the coefficients $\hat{x}$. Training 
and inference are performed using the Ridge class (with $\alpha=0$, without fitting an intercept) 
from the scikit-learn library \cite{scikit-learn}. The hyperparameters used in the models throughout 
this work are displayed in Table \ref{tab:model_hyperparameters}. Hyperparameter optimization is 
performed through Bayesian optimization using Gaussian Processes (\verb|gp_minimize)|, as implemented 
in the scikit-optimize library \cite{scikit-opt}. This is done solely on part of the training set, 
where a single snapshot of each phase is used as validation. Using these hyperparameters, we obtain 
a JLCDM with 1928 features, a model for the $\hat{n}_{00}^{00}$ components with 13 features, a model 
for the $\hat{n}^{LM}_{L0}$ components with 10 features, and a model for the $\hat{n}^{LM}_{l_1l_2}$ 
components with 120 features.

\begin{table}[!h]
\centering
\caption{Optimized hyperparameters and corresponding feature size for each of the
models generated.}
\label{tab:model_hyperparameters}
\begin{tabular}{lccccccc}
\hline
\multicolumn{1}{c}{Model} & Body  & $r_{\rm cut}$ & $n_{\rm max}$ & $l_{\rm max}$ & $r_{\rm min}$ & $\alpha$ & $\beta$ \\ \hline
JLCDM & 1B & 4.93 & 19 & -  & $-0.95$ & 7.35    & 7.20 \\
        & 2B & 4.93 & 11 & 8  & $ 0.00$ & 4.28    & 2.71 \\
CJLM  & 1B & 6.00 & 6 & -  & $0.00$ & 2.00 & 2.00 \\
       & 2B & 6.00 & 6 & 3 & $0.00$ & 2.00 & 2.00 \\
       & 3B & 6.00 & 4 & 3 & $0.00$ & 2.00 & 2.00 \\ \hline
\end{tabular}
\end{table}

\section*{Appendix C: Data and Code availability}

The data used to train and test the models (DFT charge density, structure files, and trained models) 
is available via Zenodo \cite{zenodo}. Scripts and related code for calculating the Jacobi-Legendre 
grid-based linear expansion are available at \url{https://github.com/StefanoSanvitoGroup/MLdensity}.

\end{document}